\documentclass[superscriptaddress,aps,letterpaper,10pt,twocolumn,floats,showpacs,amsmath,amsfonts,amssymb,prb]{revtex4-2}

\usepackage{graphicx}
\usepackage[colorlinks=true,citecolor=blue]{hyperref}
\usepackage[caption=false]{subfig}
\usepackage{pstricks}
\usepackage{pst-node}
\usepackage{amsmath}
\usepackage{amsbsy}
\usepackage{verbatim}
\usepackage{dsfont}
\usepackage{enumerate}
\usepackage[utf8x]{inputenc}

\begin{document}


\title{Non-Markovian thermal operations boosting the performance of quantum heat engines}

\author{Krzysztof Ptaszy\'{n}ski}
\affiliation{Institute of Molecular Physics, Polish Academy of Sciences, Mariana Smoluchowskiego 17, 60-179 Pozna\'{n}, Poland}
\email{krzysztof.ptaszynski@ifmpan.poznan.pl}

\date{\today}

\begin{abstract}
It is investigated whether non-Markovianity, i.e., the memory effects resulting from the coupling of the system to its environment, can be beneficial for the performance of quantum heat engines. Specifically, two physical models are considered. The first one is a well known single-qubit Otto engine; the non-Markovian behaviour is there implemented by replacing standard thermalization strokes with so-called extremal thermal operations which cannot be realized without the memory effects. The second one is a three-stroke engine in which the cycle consists of two extremal thermal operations and a single qubit rotation. It is shown that the non-Markovian Otto engine can generate more work-per-cycle for a given efficiency than its Markovian counterpart, whereas performance of both setups is superior to the three-stroke engine. Furthermore, both the non-Markovian Otto engine and the three-stroke engine can reduce the work fluctuations in comparison with the Markovian Otto engine, with their relative advantage depending on the performance target. This demonstrates the beneficial influence of non-Markovianity on both the average performance and the stability of operation of quantum heat engines.
\end{abstract}

\maketitle

\section{Introduction}
In recent years much research interest has been devoted to nanoscopic quantum heat engines converting heat into work (or conversely, using work to cool the environment)~\cite{quan2007, kosloff2014, myers2022}. On the fundamental level, such devices enable one to investigate basic aspects of heat-to-work conversion at the nanoscale. Studies in this area focused on topics such as fundamental resource-theoretic limits to the work extraction~\cite{skrzypczyk2014, allahverdayn2004}, engineering of optimal driving protocols~\cite{skrzypczyk2014, ma2018, erdman2019, baumer2019}, geometric approaches to thermodynamics~\cite{raz2016, brandner2020, abiuso2020} and the role of quantum coherence~\cite{feldmann2003, scully2010, scully2011, harbola2012, brandner2015, brandner2016, karimi2016, chen2017, hardal2017, brandner2017, roulet2017, bengtsson2018, ptaszynski2018, klatzow2019, holubec2019, gonzalez2019, holubec2019, rignon2021, liu2021}, relativist effects~\cite{myers2021} or cooperative coupling to the baths~\cite{mu2017}. Apart from characterizing the mean behavior of such devices, fluctuations of thermodynamic quantities, which are non-negligible at the nanoscale, recently received much attention~\cite{verley2014a, verley2014b, proesmans2015, park2016, holubec2017, holubec2018, ito2019, manikandan2019, vroylandt2020, saryal2021, holubec2021, xu2021, watanabe2022}. In particular, topics such as fluctuation theorems, characterizing the universal properties of fluctuations~\cite{sinitsyn11, campisi2014}, or thermodynamic trade-offs between power, efficiency and stability of operation on nanoengines~\cite{holubec2014, pietzonka2018, ptaszynski2018, rignon2021, liu2021} have been investigated. Though most studies so far have been theoretical in nature, quantum heat engines have been experimentally realized (or at least emulated) using setups such as trapped ions~\cite{rossnagel2016, lindenfels2019}, NV centers~\cite{klatzow2019} or spin systems~\cite{peterson2019}. Furthermore, practical applications of such engines for cooling of qubits employed in the quantum computing have been proposed~\cite{alhambra2019, campisi2022}.

This paper investigates whether the performance of nanoscopic heat engines can be improved by using non-Markovian effects. Non-Markovianity is a term used to describe a situation when dynamics of the system is not wholly determined by its actual state, but rather is influenced by the memory about its past evolution, which is encoded in the correlations between the system and its environment~\cite{breuer2002, rivas2014}. This is usually (though not strictly~\cite{wenderoth2021}) related to the strong system-environment coupling~\cite{breuer2002}. As already demonstrated, non-Markovianity is not only interesting from the fundamental point of view -- being a feature of many realistic systems --  but may also affect the performance of practical appliances, especially nanodevices. In particular, non-Markovian effects were found to be advantageous in applications such as quantum communication channels~\cite{vasile2011, bylicka2014}, quantum metrology~\cite{chin2012, zhang2017}, entanglement generation~\cite{huelga2012, heineken2021}, algorithmic cooling of qubits~\cite{alhambra2019} or photochemical switches~\cite{spaventa2022}.

In the context of quantum heat engines, the beneficial influence of non-Markovian effects on the performance of the quantum Otto engine based on a single harmonic oscillator has been previously claimed in Ref.~\cite{zhang2014}. The observed power enhancement was related to generation of states with occupation numbers exceeding the thermal value due to strong coupling to the environment; it was also claimed that such an engine can surpass the Carnot efficiency. However, subsequent works demonstrated this apparent violation of the second law of thermodynamics to be a result of neglect of the work performed during the coupling and the uncoupling of the system to the thermal bath~\cite{thomas2018, shirai2021, wiedmann2020}. When this work is taken into account, the validity of the Carnot bound is secured. Indeed, after proper inclusion of the coupling/uncoupling work, the strong coupling has been demonstrated to be detrimental rather than beneficial, though its influence can be mitigated by applying appropriate driving protocols~\cite{shirai2021, wiedmann2020}. Two other works~\cite{abiuso2019, camati2020} demonstrated the power enhancement in the non-Markovian regime for the Otto engine with a specific type of system-bath interaction in which the system is coupled to a Markovian bath through an auxiliary qubit. However, since in both works the Markovian relaxation rate was kept fixed, while coupling to the auxiliary qubit has been tuned, the observed power enhancement was actually a result of a relatively trivial enhancement of the effective system-bath coupling. Therefore, according to my best knowledge, a genuine advantage of non-Markovianity in the context of quantum heat engines so far has not been shown.

The goal of this paper is to rigorously demonstrate the beneficial influence of non-Markovian effects on the performance of a quantum heat engine based on a single qubit, without referring to any particular model of the system-bath interaction. In order to do so, I will apply resource theory methods which provide fundamental constraints on transformations between quantum states due to interaction with the thermal environment (independent of details of microscopic dynamics)~\cite{lostaglio2019, skrzypczyk2014, horodecki2013, brandao2013, brandao2015, goold2016}. This formalism has been previously used, for example, to demonstrate the enhancement of photoisomerization yield through non-Markovian effects beyond values achievable in the Markovian regime~\cite{spaventa2022}. Within this framework the state transformations due to coupling to the heat bath are described by means of so-called thermal operations, i.e., energy preserving unitary operations $U$ acting on the system and the bath; the condition of energy preservation is given by the equation ${[U,H_S+H_B]}=0$, where $H_S$ and $H_B$ are the Hamiltonians of the system and the bath, respectively. The state of the system is transformed as
\begin{align} \label{thermop}
	\rho_S \xrightarrow{TO} \mathcal{E} \rho_S = \text{Tr}_B \left[ U \rho_S \otimes \rho_B^\text{th} U^\dagger \right],
\end{align}
where $\mathcal{E}$ is the dynamical map representing the transformation, $\text{Tr}_B$ is the partial trace over the states of the bath and $\rho_B^\text{th}=e^{-\beta H_B}/\text{Tr} \left (e^{-\beta H_B} \right)$ is the Gibbs state of the bath, with $\beta=1/(k_B T)$ being the inverse temperature. It can be noted that besides its generality the applied formalism has another advantage: the energy preservation provides that the energy change of the system due to a thermal operation can be unambiguously identified with the heat exchanged with the environment (in contrast to Refs.~\cite{zhang2014, thomas2018, shirai2021, wiedmann2020}), which significantly simplifies the analysis.

The paper is organized as follows. Sec.~\ref{sec:therm} presents basic elements of the resource-theoretic framework, focusing on the distinction between Markovian and non-Markovian operations. Sec.~\ref{sec:work} analyzes the average performance of two heat engine models based on a single qubit, with heat strokes described by means of thermal operations: a well known quantum Otto engine~\cite{feldmann2000} and a much less known three-stroke engine~\cite{lobejko2020}. Sec.~\ref{sec:fluct} is concerned with the behavior of work fluctuations. Sec.~\ref{sec:exp} discusses the experimental realizability of quantum heat engines employing non-Markovian thermal operations. Finally, Sec.~\ref{sec:concl} brings the conclusions.

\section{Thermal operations} \label{sec:therm}
\subsection{Criteria of Markovianity of thermal operations}
Let us start our discussion with a rigorous definition of the distinction between Markovian and non-Markovian state transformations. The thermal operation is deemed to be Markovian when it can be expressed as the time ordered exponential~\cite{lostaglio2021}
\begin{align}
	\mathcal{E} = \mathcal{T} \left\{e^{\int_0^t \mathcal{L}_\tau d\tau} \right\}
\end{align}
of a (possibly time-dependent) Lindblad generator $\mathcal{L}_t$ defined as
\begin{align}
	\mathcal{L}_t \rho_S(t)=\mathcal{H} \rho_S(t)+\mathcal{D}_t \rho_S(t),
\end{align}
where $\mathcal{H}$ and $\mathcal{D}_t$ are generators of the unitary and the dissipative dynamics, respectively. They are given by the equations
\begin{align}
\mathcal{H} \rho_S(t) &=-i \left[H_S, \rho_S(t) \right], \\
	\mathcal{D}_t \rho_S(t) &=\sum_i r_i(t) \left[L_i \rho_S(t) L_i^\dagger- \frac{1}{2}\left\{ L_i^\dagger L_i, \rho_S(t)\right\} \right],
\end{align}
where $L_i$ and $r_i(t)$ are the Lindblad operators and the transition rates, respectively. Here and from hereon $\hbar=1$ is taken. The Lindblad generator is required to fulfill two properties~\cite{lostaglio2021}:
\begin{itemize}
	\item Stationary thermal state: the Gibbs state of the system,
	\begin{align}
\rho_S^\text{th}=\frac{e^{-\beta H_S}}{\text{Tr} \left (e^{-\beta H_S} \right)},
	\end{align}
is the stationary state of the dynamics at all times, i.e.,
\begin{align}
	\forall t: \quad \mathcal{L}_t \rho_S^\text{th}=0.
\end{align}
	\item Covariance: $\mathcal{H}$ and $\mathcal{D}_t$ commute for an arbitrary density matrix $\rho$, i.e.,
	\begin{align}
	\forall \rho: \quad	\mathcal{H} \mathcal{D}_t \rho = \mathcal{D}_t \mathcal{H} \rho.
	\end{align}
\end{itemize}

As shown by Lostaglio and Korzekwa~\cite{lostaglio2021}, any Markovian thermal operation on a $n$-level quantum system can be realized by a sequence of full or partial thermalizations acting on pairs of energy levels. Thus, in particular, a Markovian thermal operation acting an a two-level system (qubit) corresponds to its full or partial thermalization.

\subsection{Thermal operations on a single qubit}
The paper focuses on the case when the heat engine is based on a single qubit, for which an exact analytic form of arbitrary thermal operation can be found. The qubit is described by the Hamiltonian
\begin{align}
	H_S=\omega |e \rangle \langle e|,
\end{align}
where $\omega$ is the qubit energy gap and $|e \rangle$ denotes the excited state of the qubit.

In general, a dynamical map describing the evolution of an open quantum system may couple the dynamics of the diagonal elements of the density matrix (populations) and the off-diagonal ones (coherences). However, due to energy conservation, for thermal operations the evolution of populations and coherences is decoupled in the eigenstate basis, and thus may be considered in a separate way~\cite{cwiklinski2015}. Let us first consider evolution of the population vector $\mathbf{p}=(p_g,p_e)^T$, where $p_g$ and $p_e=1-p_g$ are populations of the ground and the excited state, respectively. It undergoes the transformation~\cite{lostaglio2018, lostaglio2019}
\begin{align}
	\mathbf{p} \xrightarrow{TO} \Lambda \mathbf{p},
\end{align}
where
\begin{align} \label{gstochmap}
	\Lambda(\omega,\beta,\lambda) = (1-\lambda) \begin{pmatrix} 1 & 0 \\ 0 & 1 \end{pmatrix}+\lambda \begin{pmatrix} 1-e^{-\beta \omega} & 1 \\ e^{-\beta \omega} & 0 \end{pmatrix},
\end{align}
is referred to as a Gibbs-stochastic matrix~\cite{lostaglio2018, lostaglio2019}, with parameter $\lambda \in [0,1]$ describing the strength of interaction.
 
The qubit coherences $\rho_{ge}$ and $\rho_{eg}=\rho_{ge}^*$ undergo the transformation~\cite{cwiklinski2015, lostaglio2019}
\begin{align}
	\rho_{ge} \xrightarrow{TO} \gamma \rho_{ge},
\end{align}
where $\gamma \in[0, \sqrt{1-\lambda e^{-\beta \omega}(1-\lambda)}] \leq 1$. For the sake of simplicity, in this paper the dynamics of coherences will be neglected. This corresponds to the case when $\gamma=0$, and thus the qubit is fully dephased at the end of each thermal operation.

It appears that the parameter $\lambda$ determines whether the thermal operation is Markovian or not. It is Markovian when~\cite{alhambra2019}
\begin{align}
	\lambda \leq p_g^\text{th}=\frac{1}{1+e^{-\beta \omega}},
\end{align}
where $p_g^\text{th}$ is the thermal population of the ground state corresponding to the inverse bath temperature $\beta$. In particular, $\lambda=p_g^\text{th}$ and $\lambda < p_g^\text{th}$ correspond to the full and the partial thermalization, respectively. 

\subsection{Extremal thermal operation (ETO)} \label{sec:eto}
%
\begin{figure}
	\centering
	\includegraphics[width=0.90\linewidth]{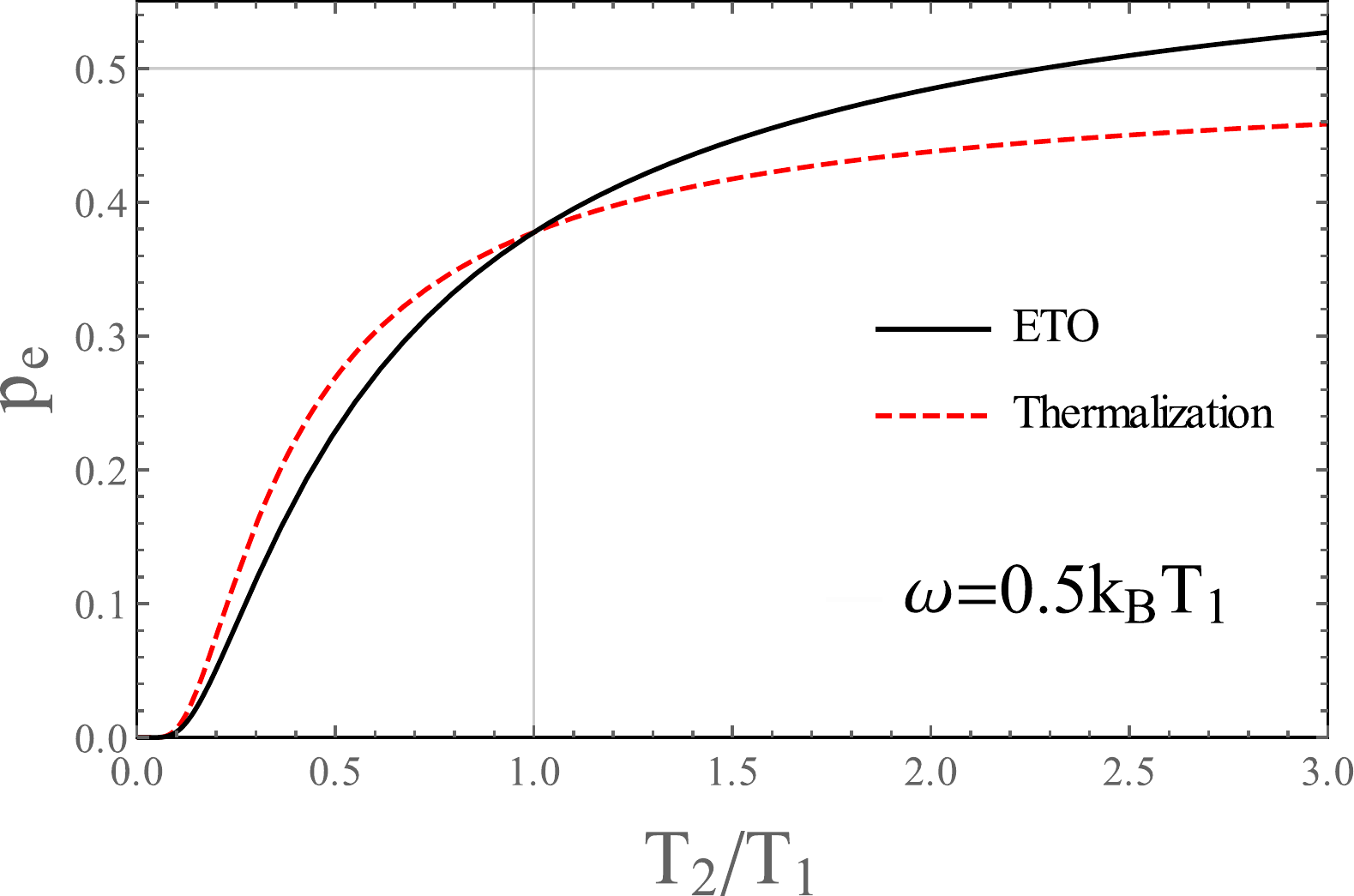}		
	\caption{The population of the excited state $p_e$ after the application of ETO (black solid line) or thermalization (red dashed line) for a qubit initialized in the thermal state with the temperature $T_1$ and the applied thermal operation corresponding to the reservoir temperature $T_2$. The qubit gap fixed at $\omega=0.5 k_B T_1$.}
	\label{fig:eto}
\end{figure}
%
Let us now consider the non-Markovian regime of $\lambda>p_g^\text{th}$, focusing on the limiting case of $\lambda=1$. The thermal operation $\Lambda$ corresponding to $\lambda=1$ is referred to as the \textit{extremal thermal operation} (ETO)~\cite{lobejko2020}. The physical nature of such an operation can be understood qualitatively by analyzing the example of a qubit initialized in the thermal state with some temperature $T_1$ and than transformed by ETO corresponding to the reservoir temperature $T_2$. Fig.~\ref{fig:eto} presents the population of the excited state $p_e$ after the application of ETO (black solid line) and after thermalization to the temperature $T_2$ (red dashed line). As one can observe, for $T_2<T_1$ ETO ``cools the qubit more'' than thermalization, i.e., population of the excited state is reduced below the thermal population corresponding to the temperature $T_2$ (and, consequently, more heat is extracted from the system than during thermalization). Conversely, for $T_2>T_1$ ETO generates the population of the excited state exceeding the thermal value. Most notably, for a high enough $T_2$ the extremal thermal operation may even produce population inversion $p_e>1/2>p_g$; this is impossible to achieve using Markovian thermal operations acting on a non-inverted state.

\section{Average work} \label{sec:work}
In this section the average work generated during a single cycle of the considered quantum heat engines is analyzed. Specifically, Secs.~\ref{sec:otto} and~\ref{sec:3st} present the models of the Otto engine and the three-stroke engine as well as analytic formulas describing work, heat and efficiency, whereas in Sec.~\ref{sec:workcomp} the performance of both models is compared.

\subsection{Otto engine} \label{sec:otto}
%
\begin{figure}
	\centering
	\includegraphics[width=0.90\linewidth]{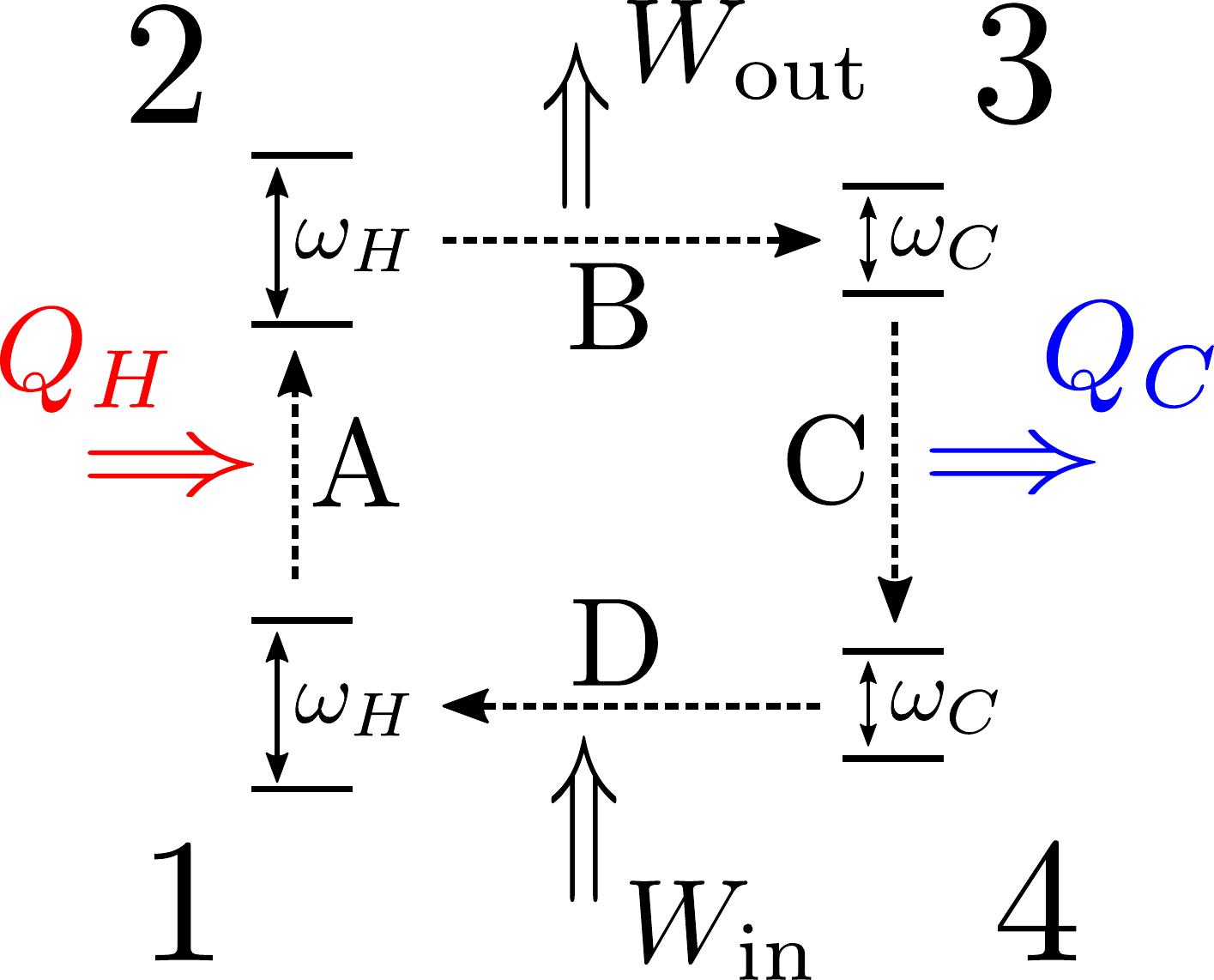}		
	\caption{A schematic representation of the Otto engine cycle. Successive strokes denoted as A, B, C, D and their starting points as 1, 2, 3, 4. $Q_H$ and $Q_C$ denote the heat flows whereas $W_\text{out}$ and $W_\text{in}$ -- the work output and the work input, respectively.}
	\label{fig:otto}
\end{figure}
%
Let me first present a principle of operation of the quantum Otto engine. It is based on a single qubit attached to two thermal baths: the hot bath $H$ with the temperature $T_H$ and the cold one $C$ with the temperature $T_C$. The operation of the Otto engine is based on a periodic coupling (decoupling) of the qubit to (from) the thermal baths, accompanied by a periodic modulation of the qubit gap $\omega$ between two values: $\omega_H$ and $\omega_C$. A schematic representation of the Otto engine cycle is presented in Fig.~\ref{fig:otto}, with strokes denoted with letters A, B, C, D and their starting points denoted with numbers 1, 2, 3, 4. The strokes can be described as follows:
\begin{enumerate}[A.]
	\item Heating: the qubit gap $\omega=\omega_H$ is taken and the system is placed into contact with the hot bath; as a result, the population of the excited state is increased from $p_{e,1}$ to $p_{e,2}$ and heat $Q_H={\omega_H (p_{e,2}-p_{e,1})}$ is delivered from the bath to the system.
	\item Work output: the qubit gap is shifted from $\omega_H$ to $\omega_C$ without change of state of the qubit ($p_{e,3}=p_{e,2}$); as a result, work $W_\text{out}={(\omega_H-\omega_C)p_{e,3}}$ is generated.
	\item Cooling: the system is placed into contact with the cold bath; as a result, the population of the excited state is decreased from $p_{e,3}$ to $p_{e,4}$ and heat $Q_C={\omega_C (p_{e,4}-p_{e,3})}$ is delivered from the system to the bath.
	\item Work input: the qubit gap is shifted from $\omega_C$ to $\omega_H$ without change of state of the qubit ($p_{e,1}=p_{e,4}$); as a result, work $W_\text{in}={(\omega_H-\omega_C)p_{e,1}}$ is delivered to the engine.
\end{enumerate}
The strokes $A$ and $C$, during which heat is exchanged with the bath, are referred to as the heat strokes, while the strokes $B$ and $D$, during which work is performed, as the work strokes.

The net work generated during a cycle is equal to 
\begin{align} \label{wotto}
	W=W_\text{out}-W_\text{in}=(\omega_H-\omega_C)(p_{e,3}-p_{e,1}),
\end{align}
and the efficiency of the Otto engine is given by a simple formula 
\begin{align} \label{efotto}
	\eta=\frac{W}{Q_H}=1-\frac{\omega_C}{\omega_H}.
\end{align}
Off course, though the equation above is mathematically well defined for arbitrary $\omega_C$ and $\omega_H$, the setup works as a heat engine only when the efficiency does not exceed the Carnot efficiency $\eta_C=1-T_C/T_H$, such that $W, Q_H>0$. 

The populations of the exited states $p_{e,1}$ and $p_{e,3}$, which are required to calculate the work-per-cycle, can be found in the following way. First, the steady state of the system in point 1 is found by solving the equation
\begin{align}
	\Lambda_C \Lambda_H \mathbf{p}_1 =\mathbf{p}_1,
\end{align}
where $\Lambda_C$ and $\Lambda_H$ are Gibbs-stochastic maps given by Eq.~\eqref{gstochmap} describing the interaction with the cold bath and the hot bath, respectively:
\begin{align} \Lambda_\alpha=\Lambda(\omega_\alpha,\beta_\alpha,\lambda_\alpha),
\end{align}
with $\alpha \in \{C,H\}$. The populations in point 3 are then calculated as
\begin{align}
	\mathbf{p}_3=\Lambda_H \mathbf{p}_1.
\end{align}

Two versions of the Otto engine are then considered: ones with Markovian and non-Markovian thermal operations. For the Markovian Otto engine I take
\begin{align}
	\lambda_\alpha=\frac{1}{1+e^{-\beta_\alpha \omega_\alpha}},
\end{align}
which corresponds to thermalization of the system. The stationary excited state populations take then the thermal values
\begin{align} \label{p1m}
	p_{e,1}^\text{M} &=\frac{1}{1+e^{\beta_C \omega_C}}, \\ \label{p3m}
	p_{e,3}^\text{M} &=\frac{1}{1+e^{\beta_H \omega_H}}.
\end{align}
For the non-Markovian setup $\lambda_\alpha=1$ is taken, which generates the extremal thermal operation. The excited state populations read then as
\begin{align} \label{p1nm}
	p_{e,1}^\text{NM} &=\frac{e^{\beta_H \omega_H}-1}{e^{\beta_C \omega_C+\beta_H \omega_H}-1}, \\ \label{p3nm}
	p_{e,3}^\text{NM} &=\frac{e^{\beta_C \omega_C}-1}{e^{\beta_C \omega_C+\beta_H \omega_H}-1}.
\end{align}

\subsection{Three-stroke engine} \label{sec:3st}
%
\begin{figure}
	\centering
	\includegraphics[width=0.90\linewidth]{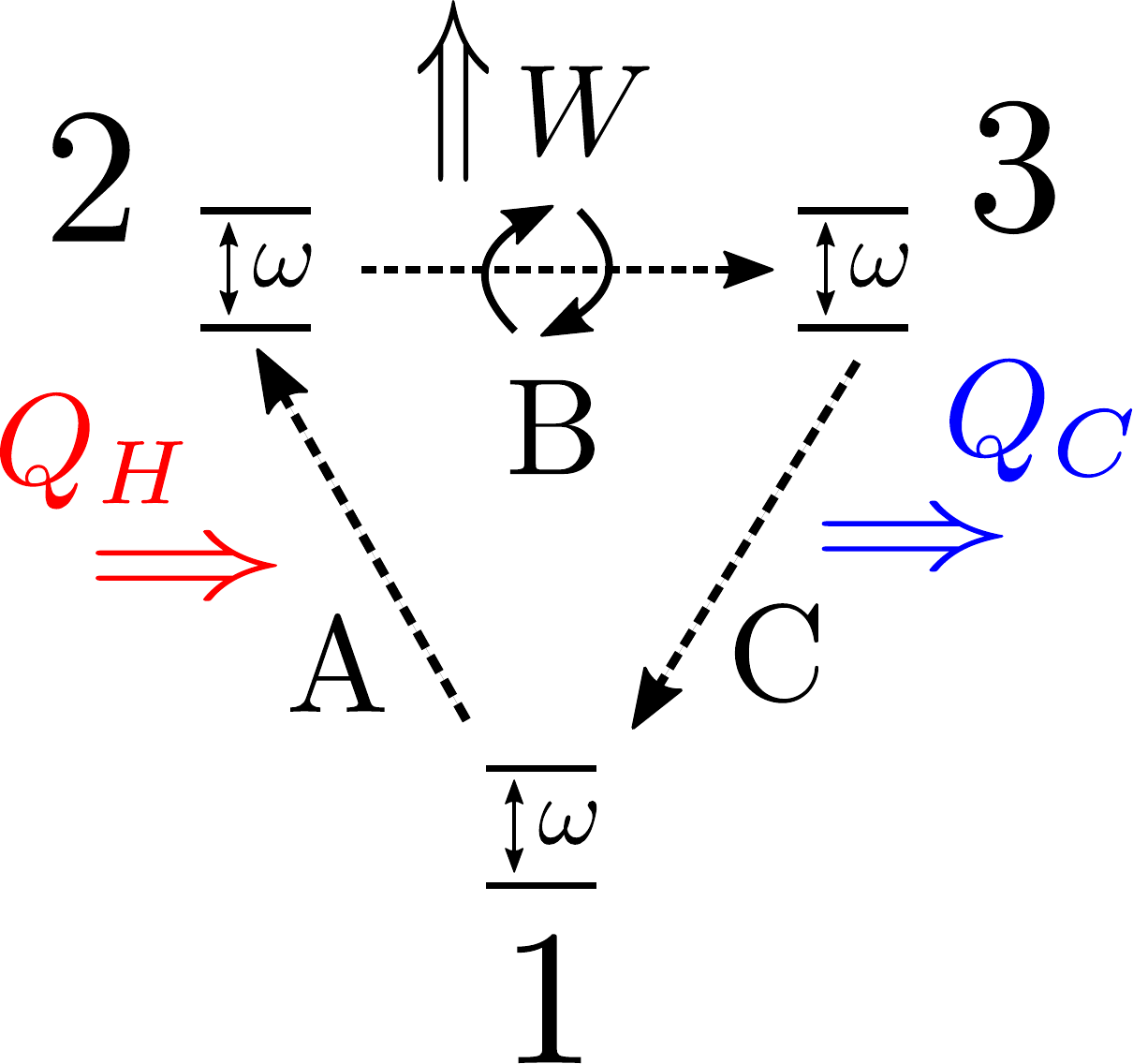}		
	\caption{A schematic representation of the three-stroke engine cycle. Successive strokes denoted as A, B, C and their starting points as 1, 2, 3. $Q_H$ and $Q_C$ denote the heat flows whereas $W$ -- the work output.}
	\label{fig:3stroke}
\end{figure}
%
Let us now turn our attention to a less known model of a quantum heat engine, namely, the three-stroke engine proposed by {\L}obejko \textit{et al.}~\cite{lobejko2020}. As before, it is based on a single qubit attached to two thermal baths. A schematic representation of its operation is presented in Fig.~\ref{fig:3stroke}, with strokes denoted as A, B, C and starting points of the strokes as 1, 2, 3. The strokes can be described as follows:
\begin{enumerate}[A.]
	\item Heating: the system is placed into contact with the hot bath; as a result, the population of the excited state is increased from $p_{e,1}$ to $p_{e,2}$ and heat $Q_H={\omega (p_{e,2}-p_{e,1})}$ is delivered from the bath to the system.
	\item Work extraction stroke: the qubit is coherently flipped such that populations of the ground and the excited states are interchanged: $p_{e,3}=p_{g,2}=1-p_{e,2}$; as a result work
		\begin{align} \label{work3st}
		W=\omega(p_{e,2}-p_{g,2})=\omega (2p_{e,2}-1)
	\end{align}
is generated.
	\item Cooling: the system is placed into contact with the cold bath; as a result, the population of the excited state is decreased from $p_{e,3}$ to $p_{e,1}$ and heat $Q_C={\omega (p_{e,1}-p_{e,3})}$ is delivered from the system to the bath.
\end{enumerate}
The be precise, in original work of {\L}obejko \textit{et al.}~\cite{lobejko2020} the work stroke consisted of the energy exchange with the battery; here it is replaced by the qubit flip for the sake of simplicity. As Eq.~\eqref{work3st} implies, the engine generates work only in the case when heating produces the population inversion: $p_{e,2}>1/2>p_{g,2}$. Therefore, since the Markovian thermal operations cannot generate the population inversion, the engine requires non-Markovian effects to operate.

Analogously to the Otto engine, the stationary populations in point 1 can be found by solving the equation
\begin{align}
	\Lambda_C \begin{pmatrix} 0 & 1 \\ 1 & 0 \end{pmatrix} \Lambda_H \mathbf{p}_1 = \mathbf{p}_1,
\end{align}
where
\begin{align}
	\Lambda_\alpha=\Lambda(\omega,\beta_\alpha,\lambda_\alpha)
\end{align}
is the Gibbs-stochastic matrix given by Eq.~\eqref{gstochmap}. Probability vectors in points 2 and 3 read as
\begin{align}
	\mathbf{p}_2 &=\Lambda_H \mathbf{p}_1, \\
	\mathbf{p}_3 &=\begin{pmatrix} 0 & 1 \\ 1 & 0 \end{pmatrix} \Lambda_H \mathbf{p}_1.
\end{align}

From hereon, the extremal thermal operations will be applied and thus $\lambda_\alpha=1$ will be taken. The stationary excited state populations read then as
\begin{align}
	p_{e,1} &=\frac{1}{1+e^{(\beta_H+\beta_C) \omega}}, \\
	p_{e,2} &=\frac{1}{e^{\beta_H \omega }+e^{-\beta_C \omega}}, \\
	p_{e,3} &=1-\frac{1}{e^{\beta_H \omega}+e^{-\beta_C \omega}}.
\end{align}

Using the expressions above, the average work-per-cycle and the efficiency are given by the equations~\cite{lobejko2020}
\begin{align}
 W &= \omega \left[ \frac{2}{e^{\beta_H \omega }+e^{-\beta_C \omega}} -1 \right], \\
 \eta &=\frac{W}{Q_H}= 1-\frac{e^{\beta_H \omega}-1}{1-e^{-\beta_C \omega}}.
\end{align}

\subsection{Comparison} \label{sec:workcomp}
I will now compare the average work-per-cycle as a function of efficiency $\eta$ for a fixed Carnot efficiency $\eta_C$. Before that, let me first note that the Otto engine and the three-stroke engine are characterized by different numbers of free parameters. The three-stroke engine is characterized by three free parameters: $\omega$, $T_H$ and $T_C$. As a result, by fixing $\eta$ and $\eta_C$ one fully determines the ratio $W/(k_B T_H)$. In contrast, the Otto engine is characterized by four free parameters: $\omega_H$, $\omega_C$, $T_H$ and $T_C$. Therefore, for a chosen efficiency $\eta$ the ratio $\omega_C/\omega_H$ becomes fixed [see Eq.~\eqref{efotto}], however, a single qubit gap $\omega_H$ can be tuned. Therefore, the presented work-per-cycle is here a maximum value, which is optimized over the full range of $\omega_H$.

%
\begin{figure}
	\centering
	\includegraphics[width=0.90\linewidth]{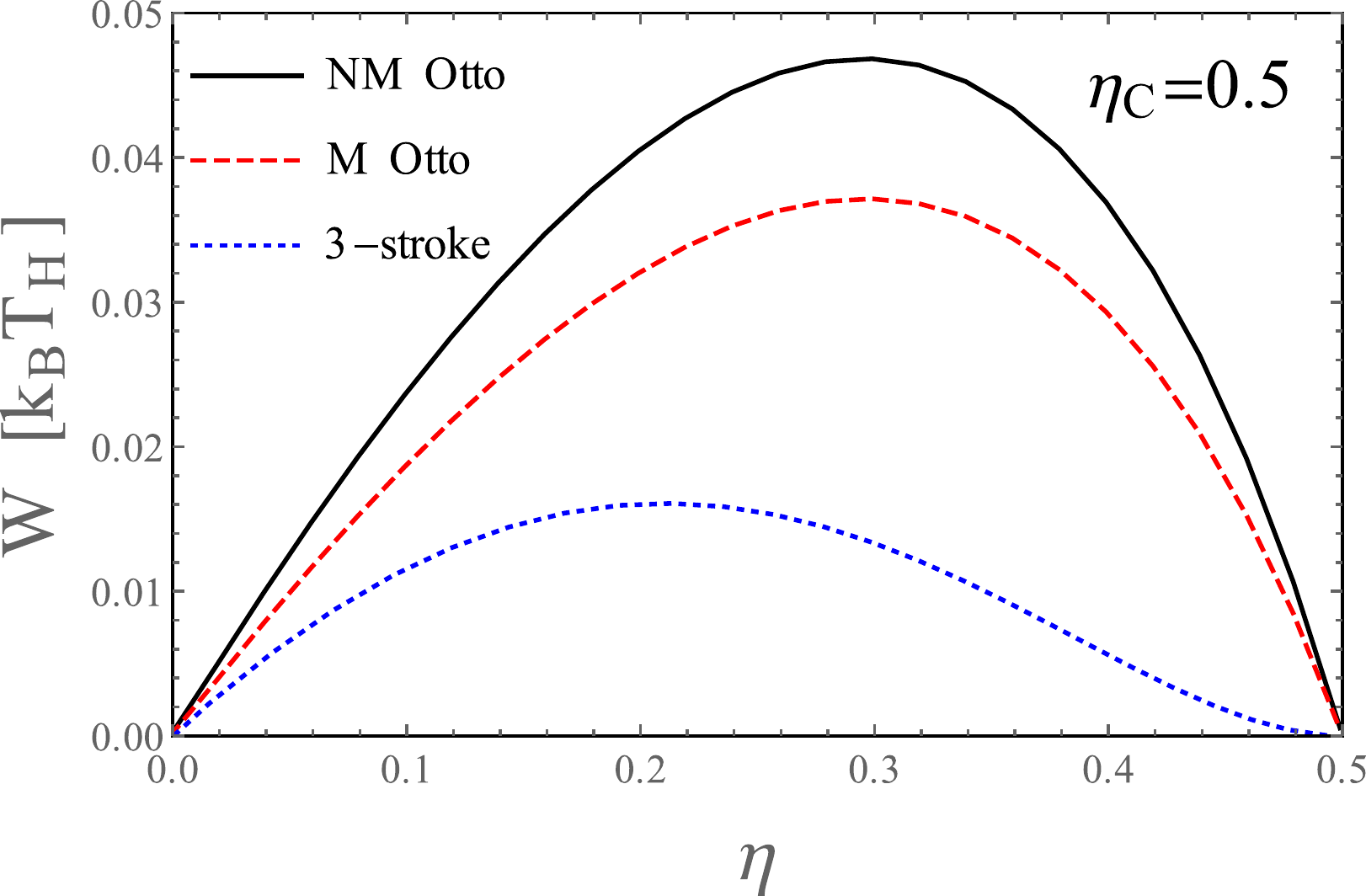}		
	\caption{The average work-per-cycle as a function of efficiency $\eta$ for the non-Markovian Otto engine (black solid line), the Markovian Otto engine (red dashed line) and the three-stroke engine (blue dotted line). The Carnot efficiency fixed at $\eta_C=1-T_C/T_H=0.5$.}
	\label{fig:work}
\end{figure}
%
The results are presented in Fig.~\ref{fig:work}. As one can observe, the non-Markovian Otto engine generates more work-per-cycle than the Markovian one. This is because by heating (cooling) the qubit through the extreme thermal operation one can produce a higher (lower) population of the excited state than using thermalization: $p_{e,3}^\text{NM}>p_{e,3}^\text{NM}$ and $p_{e,1}^\text{NM}<p_{e,1}^\text{M}$ [see Eqs.~\eqref{p1m}--\eqref{p3nm}]. Therefore, the work-per-cycle, which is proportional to the difference ${p_{e,3}-p_{e,1}}$ [see Eq.~\eqref{wotto}], is larger for the non-Markovian case. This demonstrates that non-Markovianity can be beneficial for the performance of quantum heat engines. Let me note that a similar advantage of non-Markovian thermal operations for the purpose of cooling the qubit has been previously demonstrated by Alhambra \textit{et al.}~\cite{alhambra2019}

One can also observe that in a full parameter range both the Markovian and the non-Markovian engine generate more work than the three-stroke engine. This might be related to smaller number of free parameters of the three-stroke engine, which hinders the performance optimization. Note that this conclusion differs from the one provided by {\L}obejko \textit{et al.}~\cite{lobejko2020}, who stated that the three-stroke engine becomes optimal for high efficiencies. However, in the mentioned paper the work was expressed in the units of ``energy gap'', which was defined in a different way for the model analyzed: as ${\omega_H-\omega_C}$ for the Otto engine while as $\omega$ for the three-stroke engine. In this article, in contrast, the same units of $k_B T_H$ are used for both models.

Let me finally note that the work-per-cycle may not always be an optimal measure of the average performance. Indeed, for a Markovian heat engine with a fixed maximum value of the system-bath coupling the maximum power is achieved in the limit of cycle period going to 0, in which the work-per-cycle also goes to 0~\cite{erdman2019}. However, the work-per-cycle might be still a valuable measure, for example, when maximum frequency of driving is limited due to technical constraints. 

\section{Work fluctuations} \label{sec:fluct}
So far, the paper focused on the average performance of heat engines. Let us now turn to another topic which received much attention in the recent years, namely, the work fluctuations~\cite{holubec2014, verley2014a, verley2014b, proesmans2015, park2016, holubec2017, holubec2018, pietzonka2018, ptaszynski2018, ito2019, manikandan2019, vroylandt2020, saryal2021, holubec2021, xu2021, watanabe2022, sinitsyn11, campisi2014, rignon2021, liu2021}. This field of investigation is related to the issue of stability of operation of heat engines: a good engine is expected to generate a well defined amount of work during a given time period, and thus, to minimize fluctuations. As a measure of work fluctuations this paper will use the variance of work generated during $N$ successive cycles,
\begin{align}
\langle \Delta w_N^2 \rangle =\left \langle w_N^2 \right \rangle-\left \langle w_N \right \rangle^2,
\end{align}
where $\Delta w_N=w_N-\langle w_N \rangle$. Here lowercase $w_N$ denotes the stochastic (fluctuating) work generated during $N$ cycles to avoid confusion with capital $W$ denoting the average work-per-cycle. In Sec.~\ref{sec:fcs} the methods used to calculate this quantity will be presented while Sec.~\ref{sec:fluctres} presents the results.

\subsection{Methods} \label{sec:fcs}
The work fluctuations can be determined by means of the formalism of full counting statistics~\cite{chiuchiu2018, touchette2009}. Within this framework one defines the counting-field-dependent stochastic map $\Pi_\chi$ describing the evolution of the system during a single cycle. For the Otto engine such a map reads
		\begin{align}
			\Pi_{\chi,\text{Otto}} &=B_- \Lambda_C B_+ \Lambda_H \end{align}
with		
		\begin{align}
			B_{\pm}=\begin{pmatrix} 1 & 0 \\ 0 & \exp[\pm \chi(\omega_H-\omega_C)] \end{pmatrix}.
		\end{align}
Here the counting field $\pm \chi$ counts the events during which the energy of the excited state is decreased (increased) which corresponds to the positive (negative) work generation. The analogous map for the three-stroke engine reads
		\begin{align}
			\Pi_{\chi,\text{3-stroke}} &= \Lambda_C \begin{pmatrix} 0 & \exp[\chi \omega] \\ \exp[-\chi \omega]  & 0 \end{pmatrix} \Lambda _H.
		\end{align}

The average work generated during $N$ cycles, as well as the work variance, can be then calculated using the equations	
			\begin{align} 
		\langle w_N \rangle &=\left[ \frac{d}{d\chi} G_N(\chi) \right]_{\chi=0}, \\ 
		\langle \Delta w_N^2 \rangle&=\left[ \frac{d^2}{d\chi^2} G_N(\chi) \right]_{\chi=0},
	\end{align}
where
		\begin{align} 
			G_N(\chi)=\ln \left [\mathbf{1}^T (\Pi_\chi)^N \mathbf{p}_1 \right]
		\end{align}
is the cumulant generating function. Here $\mathbf{1}=(1,1)^T$ while $\mathbf{p}_1$, as in Sec.~\ref{sec:work}, is the stationary state of the system in point 1. Since the average value of work generated in each cycle is the same, the average work generated during $N$ cycles grows linearly with the number of cycles: $\langle w_N \rangle=N W$. However, as will be later discussed more thoroughly, the higher work cumulants may not grow linearly due to correlations of work generated in the successive cycles. In particular, the work variance grows sublinearly ($\langle \Delta w_N^2 \rangle < N \langle \Delta w_1^2 \rangle$) for negative intercycle correlations while supralinearly ($\langle \Delta w_N^2 \rangle > N \langle \Delta w_1^2 \rangle$) in the opposite case~\cite{xu2021}.

To characterize the long-time properties of work fluctuations it is useful to consider the limit of infinite number of cycles ($N \rightarrow \infty$). Fluctuations in such a limit are well defined when described by means of scaled cumulants, i.e., work cumulants divided by $N$. In particular, the scaled mean work and the work variance can be calculated as~\cite{touchette2009, chiuchiu2018}
	\begin{align} 
	\lim_{N \to \infty} \frac{\langle w_N \rangle}{N} &=\left[ \frac{d}{d\chi} g(\chi) \right]_{\chi=0}, \\ \nonumber
	\lim_{N \to \infty} \frac{\langle \Delta w^2_N \rangle}{N} &=\left[ \frac{d^2}{d\chi^2} g(\chi) \right]_{\chi=0},
\end{align}
where 
\begin{align} 
	g(\chi)=\ln \lambda_0(\chi)
\end{align}
is the scaled cumulant generating function, with $\lambda_0(\chi)$ being the dominant eigenvalue of $\Pi_{\chi}$.

\subsection{Results} \label{sec:fluctres}
%
\begin{figure}
	\centering
	\includegraphics[width=0.90\linewidth]{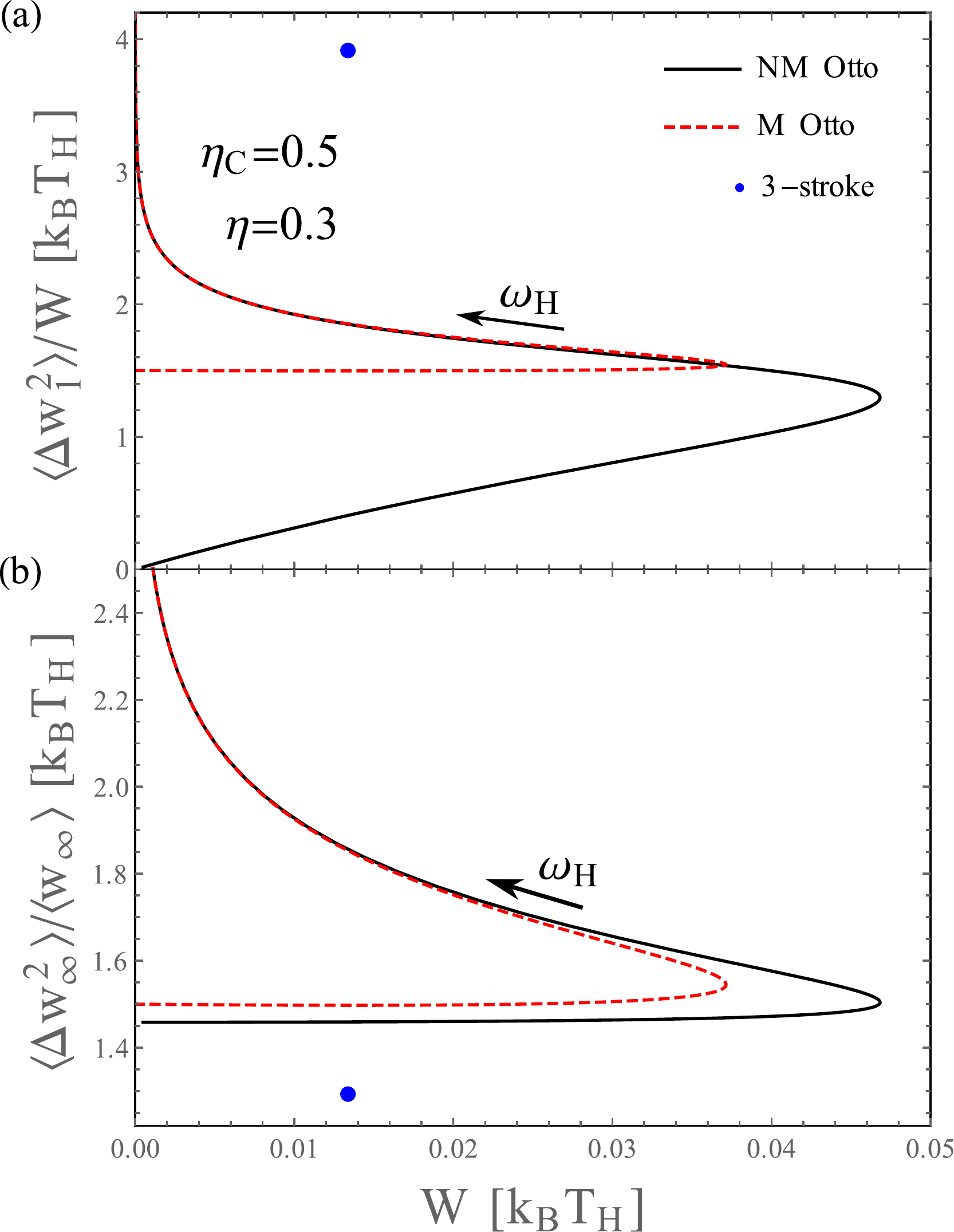}		
	\caption{The ratio of the work variance to the average work for a single-cycle (a) and in the limit of infinite number of cycles (b) for the non-Markovian Otto engine (black solid line), the Markovian Otto engine (red dashed line) and the three-stroke engine (blue point). The arrow near the curves points the direction of increasing $\omega_H$. Results for $\eta=0.3$ and $\eta_C=1-T_C/T_H=0.5$.}
	\label{fig:fluct}
\end{figure}
%
Let me now present the results. Fig.~\ref{fig:fluct} shows the ratio of the work variance to the mean work as a function of work-per-cycle for a fixed efficiency $\eta$ and the Carnot efficiency $\eta_C=1-T_C/T_H$. The fluctuations are analyzed in two opposite limits: the short-time and the long-time behavior is described by means of fluctuations during a single cycle [Fig~\ref{fig:fluct}~(a)] and in the limit of infinite number of cycles [Fig~\ref{fig:fluct}~(b)], respectively. As one can note, the work fluctuations are represented by curves for the  Otto engines while by a point for the three-stroke engine. This is related to a different number of free parameters of the models considered, which was discussed in Sec.~\ref{sec:workcomp}: for the three-stroke engine fluctuations and the average work become fixed for a specific choice of $\eta$ and $\eta_C$, whereas for the Otto engine different values can be accessed by tuning $\omega_H$; the direction of increasing $\omega_H$ is denoted by the arrow near the curves.

One can observe that for a single cycle [Fig.~\ref{fig:fluct}~(a)] the non-Markovian Otto engine generates less fluctuations for a given work-per-cycle than the Markovian one. For $\omega_H \rightarrow 0$ the ratio $\langle \Delta w_1^2 \rangle/W$ goes to 0, which is, however, also accompanied by the mean value of work going to 0. Additionally, both the non-Markovian and the Markovian Otto engine generate less fluctuations than the three-stroke engine. Interestingly, a qualitatively different behavior is observed in the limit of infinite number of cycles [Fig.~\ref{fig:fluct}~(b)]. While the non-Markovian Otto engine still generates less fluctuations than the Markovian one, the difference is much less pronounced and the ratio $\langle \Delta w_\infty^2 \rangle/\langle w_\infty \rangle$ no longer goes to 0 for $\omega_H \rightarrow 0$. Even more importantly, the three-stroke engine now generates the smallest amount of work fluctuations, which is opposite to the single-cycle case. However -- for the parameters considered -- the non-Markovian Otto engine can generate about 4 times more work-per-cycle than the three-stroke engine for the fluctuation-to-work ratio $\langle \Delta w_\infty^2 \rangle/\langle w_\infty \rangle$ larger only by about 10\%. Therefore, either three-stroke engine or the non-Markovian Otto engine may be optimal depending on whether one is more interested in maximizing power or minimizing the work fluctuations.

As shown before by Xu and Watanabe~\cite{xu2021}, a different behavior of single-cycle and long-time fluctuations is a result of correlations of the work generated in the successive cycles. Specifically, the long-time fluctuations are enhanced (reduced) by positive (negative) correlations. This can be demonstrated by analyzing the work covariance
\begin{align}
	\text{Cov}(w)=\langle \Delta w_I \Delta w_{II} \rangle,
\end{align}
where $w_I$ and $w_{II}$ are the amounts of work generated in two successive cycles. As one can easily show, the work covariance is related to work variances through the relation
\begin{align}
	\langle \Delta w_2^2 \rangle = 2 \langle \Delta w_1^2 \rangle + 2 \text{Cov}(w).
\end{align}

%
\begin{figure}
	\centering
	\includegraphics[width=0.90\linewidth]{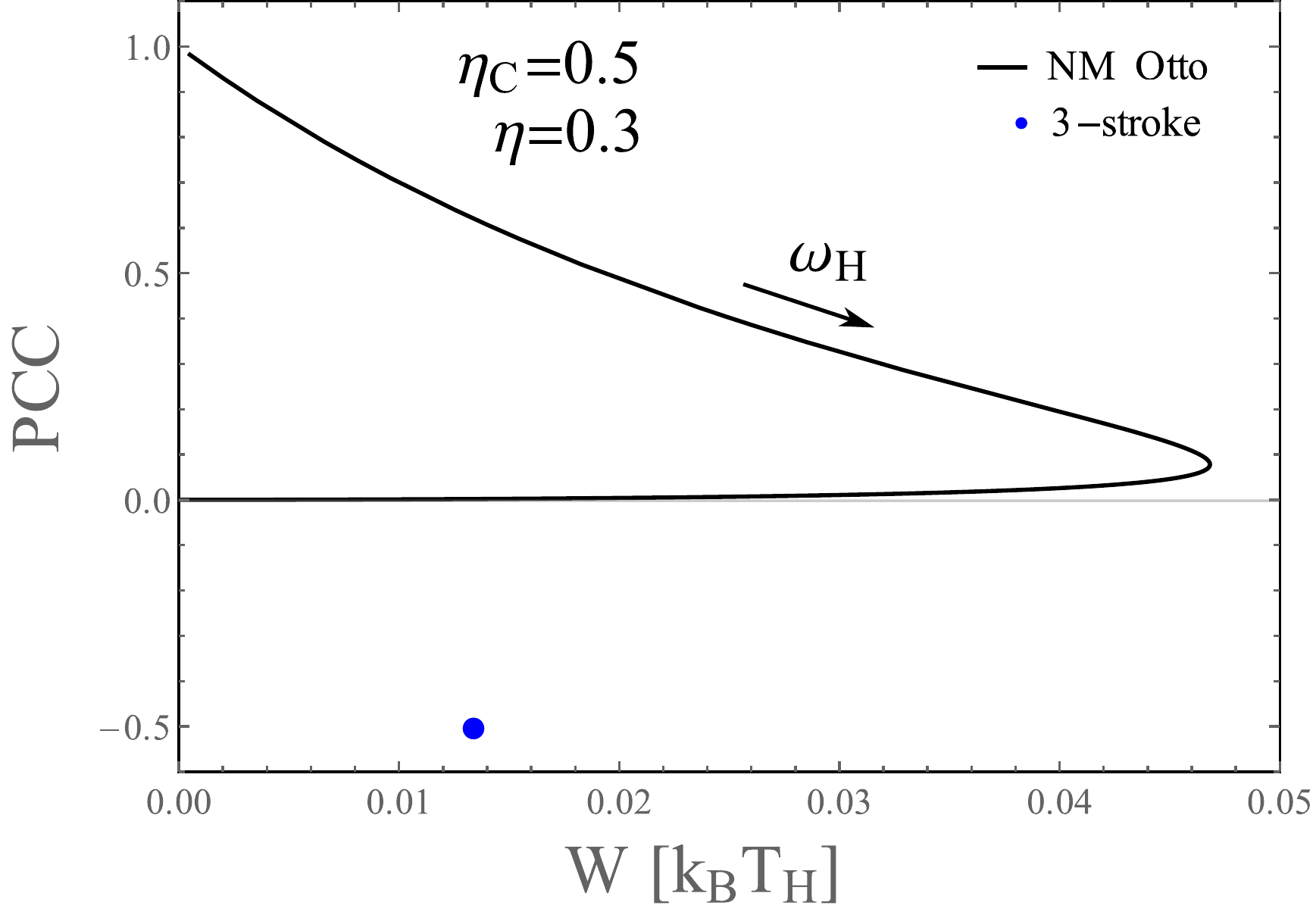}		
	\caption{The Pearson correlation coefficient of work correlations as a function of work-per-cycle for the non-Markovian Otto engine (black solid line) and the three-stroke engine (blue point). The arrow near the black curve points the direction of increasing $\omega_H$. Results for $\eta=0.3$ and $\eta_C=1-T_C/T_H=0.5$.}
	\label{fig:pearson}
\end{figure}
%
Fig.~\ref{fig:pearson} shows the Pearson correlation coefficient
\begin{align}
	\text{PCC}=\frac{\text{Cov}(w)}{\langle \Delta w_1^2 \rangle}=\frac{\langle \Delta w_2^2 \rangle}{2\langle \Delta w_1^2 \rangle}-1
\end{align}
for the non-Markovian Otto engine and the three-stroke engine, with parameters as in Fig.~\ref{fig:fluct}. It takes values within the range $\text{PCC} \in [-1,1]$, with a positive (negative) sign corresponding to positive (negative) correlations. The intercycle correlations are not present for the Markovian Otto engine because the system is resetted to the thermal state after each heat stroke, and thus the initial state at the beginning of each cycle does not depend on the previous evolution. As one can observe, the non-Markovian Otto regime is characterized by positive intercycle work correlations which leads to the enhancement of long-time fluctuations. In particular, perfect positive correlations ($\text{PCC}=1$) are observed in the limit $\omega_H \rightarrow 0$. Conversely, the work correlations are negative for the three-stroke engine. Interestingly, the Pearson correlation coefficient is then given by a simple formula
\begin{align}
\text{PCC}_\text{3-stroke} = -e^{-(\beta_C+\beta_H) \omega}
\end{align}
which implies that the work generated in the successive cycles becomes perfectly anti-correlated ($\text{PCC}=-1$) for $\omega \rightarrow 0$.

\section{Experimental realizability} \label{sec:exp}
Finally, let us discuss how heat engines with non-Markovian thermal operations can be realized experimentally. First, it was shown that the extremal thermal operation described in Sec.~\ref{sec:eto} can be realized by coupling the qubit to the thermal state of a single bosonic mode
\begin{align}
	\rho_\text{bos}^\text{th} = \frac{e^{-\beta \Omega a^\dagger a}}{\text{Tr} \left( e^{-\beta \Omega a^\dagger a} \right)},
\end{align}
with the excitation energy $\Omega$ resonant with the qubit gap $\omega$, i.e., $\Omega=\omega$; here $a^\dagger$ and $a$ are the bosonic creation and annihilation operators, respectively. Then, one needs to perform the energy-preserving unitary operation~\cite{lostaglio2018}
\begin{align}
	U=|g,0 \rangle \langle g,0|+\sum_{n=1}^\infty \left(|g,n \rangle \langle e,n-1|+|e,n-1 \rangle \langle g,n| \right),
\end{align}
where $g$ ($e$) denotes the ground (excited) state of the qubit and $n$ is the number of excitations of the bosonic mode. Such an operation exchanges a single excitation between the qubit and the bosonic mode. It can be exactly realized by means of the intensity-dependent Jaynes-Cummings interaction Hamiltonian~\cite{naderi2005, aberg2014}
\begin{align}
 \tilde{H}_{JC}=J \left[\sigma_+ (a a^\dagger)^{-1/2} a+ \sigma_- (a a^\dagger)^{-1/2} a^\dagger \right],
\end{align}
where $J$ is the coupling strength while $\sigma_+=|e \rangle \langle g|$ and $\sigma_-=|g \rangle \langle e|$ are raising and lowering operators, respectively. Bera \textit{et al.}~\cite{bera2022} shown that such a Hamiltonian can be realized by adding some engineered anharmonicity to the bosonic mode. Alternatively, as shown by Alhambra \textit{et al.}~\cite{alhambra2019}, a good approximation of the extremal thermal operation can be realized by using the standard Jaynes-Cummings interaction Hamiltonian
\begin{align}
	H_{JC}=J \left(\sigma_+ a + \sigma_- a^\dagger \right).
\end{align}

%
\begin{figure}
	\centering
	\includegraphics[width=0.90\linewidth]{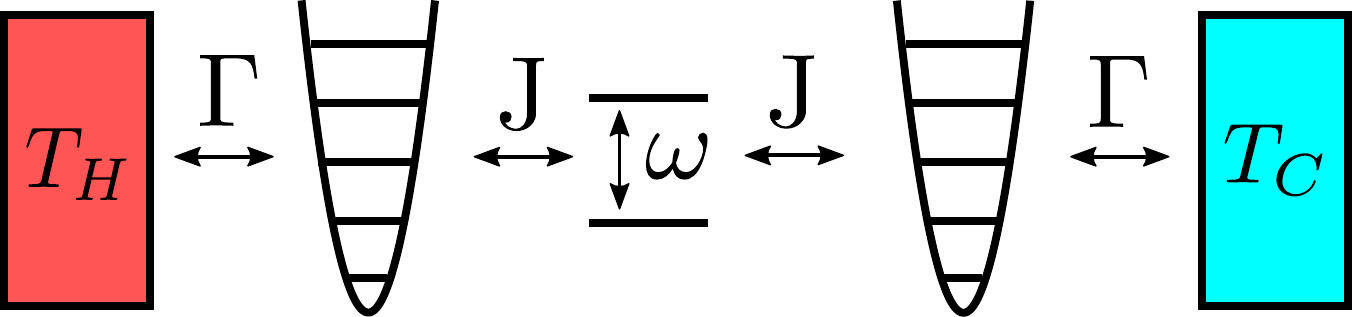}		
	\caption{A schematic idea of experimental realization of the non-Markovian heat engine. A qubit is coupled to Markovian thermal baths with temperatures $T_H$ and $T_C$ via auxiliary bosonic modes. Here $J$ denotes (possibly tunable) Jaynes-Cummings coupling between the qubit and bosonic modes, while $\Gamma$ is the relaxation rate of the bosonic mode.}
	\label{fig:exp}
\end{figure}
%
To realize a continuous work generation, the bosonic modes need to be either resetted to the thermal state after each heat stroke or replaced with another one. The former option can be implemented by coupling the bosonic levels to Markovian thermal baths. Therefore, finally, the non-Markovian heat engine can be realized by using a qubit coupled to Markovian thermal baths through auxiliary bosonic modes (Fig.~\ref{fig:exp}). In general, the coupling to auxiliary bosons needs to be tunable to enable switching of the system-bath coupling; however, for the Otto engine the effective coupling to the bosonic modes is modulated just by tuning the qubit gap $\omega$, since the excitation exchange is hampered in the non-resonant regime ($\omega \neq \Omega$)~\cite{guthrie2021}. An experimental realization of such a setup may be based, for example, on superconducting circuits with LC resonators playing the role of bosonic modes and the attached resistors playing the role of Markovian baths~\cite{ronzani2018, senior2020}; indeed, the Otto engine based on such an architecture has been previously theoretically proposed~\cite{karimi2016} and experimental steps towards its realization has been made~\cite{guthrie2021}. Another physical realization may be based on atoms in optical cavities~\cite{bera2022}.

\section{Conclusions} \label{sec:concl}
The work investigated the performance of two types of a quantum engine based on a single qubit: the Otto engine and the three-stroke engine. Interaction with the thermal baths have been described within the resource-theoretic framework by means of thermal operations, i.e., the energy-preserving unitaries acting on the system and its environment. Within this approach the fundamental limitations of heat engines employing the heat strokes based on Markovian thermalization could have been analyzed. It was shown that the Otto engine employing the non-Markovian dynamics can generate more work during a single cycle then its Markovian counterpart; at the same time, both setups produce more work then the three-stroke engine, which can operate only in the non-Markovian regime. It was also demonstrated that the non-Markovian Otto engine can generate less work fluctuations than the Markovian one. Furthermore, even less fluctuations are produced by the three-stroke engine, albeit only for a low power; therefore, the relative advantage of the non-Markovian Otto engine and the three-stroke engine depends on whether the maximization of power or minimization of fluctuations is desired. Additionally, the reduction of work fluctuations in the three-stroke engine have been demonstrated to be a result of correlations of the work generated in the successive cycles; a similar behavior has been investigated previously in the classical case~\cite{xu2021}.

On a more general level, in the spirit of some previous works~\cite{alhambra2019, spaventa2022}, the paper confirms the particular usefulness of resource-theoretic methods for demonstrating a genuine beneficial influence of non-Markovianity on the performance of quantum nanodevices. The main advantage of this approach is its generality: it provides fundamental constraints on Markovian evolution, without assuming any specific model of microscopic dynamics. This may inspire future studies on applications of non-Markovianity in other areas of quantum technology.

\begin{acknowledgments}
The author has been supported by the National Science Centre, Poland, under the project No.~2017/27/N/ST3/01604, and by the Scholarships of Minister of Science and Higher Education.
\end{acknowledgments}


\begin{thebibliography}{}
\bibitem{quan2007}
H. T. Quan, Y.-X. Liu, C. P. Sun, and F. Nori, Quantum thermodynamic cycles and quantum heat engines, \href{\doibase 10.1103/PhysRevE.76.031105}{Phys. Rev. E \textbf{76}, 031105 (2007)}.

\bibitem{kosloff2014}
R. Kosloff and A. Levy, Quantum Heat Engines and Refrigerators: Continuous Devices, \href{https://doi.org/10.1146/annurev-physchem-040513-103724}{Annu. Rev. Phys. Chem. \textbf{65}, 365 (2014)}.

\bibitem{myers2022}
N. M. Myers, O. Abah, and S. Deffner, Quantum thermodynamic devices: from theoretical proposals to experimental reality, \href{https://doi.org/10.48550/arXiv.2201.01740}{arXiv:2201.01740 (2022)}.

\bibitem{allahverdayn2004}
A. E. Allahverdyan, R. Balian, and T. M. Nieuwenhuizen, Maximal work extraction from finite quantum systems, \href{\doibase 10.1209/epl/i2004-10101-2}{Europhys. Lett. \textbf{67}, 565 (2004)}.

\bibitem{skrzypczyk2014}
P. Skrzypczyk, A. J. Short, and S. Popescu, Work extraction and thermodynamics for individual quantum systems, \href{\doibase 10.1038/ncomms5185}{Nat. Commun. \textbf{5}, 4185 (2014)}.

\bibitem{ma2018}
Y.-H. Ma, D. Xu, H. Dong, and C.-P. Sun, Optimal operating protocol to achieve efficiency at maximum power of heat engines, \href{\doibase 10.1103/PhysRevE.98.022133}{Phys. Rev. E \textbf{98}, 022133 (2018)}.

\bibitem{erdman2019}
P. A Erdman, V. Cavina, R. Fazio, F. Taddei, and V. Giovannetti, Maximum power and corresponding efficiency for two-level heat engines and refrigerators: optimality of fast cycles, \href{\doibase 10.1088/1367-2630/ab4dca}{New J. Phys. \textbf{21}, 103049 (2019)}.

\bibitem{baumer2019}
E. B\"{a}umer, M. Perarnau-Llobet, P. Kammerlander, H. Wilming, and R. Renner, Imperfect Thermalizations Allow for Optimal Thermodynamic Processes, \href{\doibase 10.22331/q-2019-06-24-153}{Quantum \textbf{3}, 153 (2019)}.

\bibitem{raz2016}
O. Raz, Y. Suba\c{s}{\i}, and R. Pugatch, Geometric Heat Engines Featuring Power that Grows with Efficiency, \href{\doibase 10.1103/PhysRevLett.116.160601}{Phys. Rev. Lett. \textbf{116}, 160601 (2016)}.

\bibitem{brandner2020}
K. Brandner and K. Saito, Thermodynamic Geometry of Microscopic Heat Engines, \href{\doibase 10.1103/PhysRevLett.124.040602}{Phys. Rev. Lett. \textbf{124}, 040602 (2020)}.

\bibitem{abiuso2020}
P. Abiuso, H. J. D. Miller, M. Perarnau-Llobet, and M. Scandi, Geometric Optimisation of Quantum Thermodynamic Processes, \href{\doibase 10.3390/e22101076}{Entropy \textbf{22}, 1076 (2020)}.

\bibitem{feldmann2003}
T. Feldmann and R. Kosloff, Quantum four-stroke heat engine: Thermodynamic observables in a model with intrinsic friction, \href{\doibase 10.1103/PhysRevE.68.016101}{Phys. Rev. E \textbf{68}, 016101 (2003)}.

\bibitem{scully2010}
M. O. Scully, Quantum Photocell: Using Quantum Coherence to Reduce Radiative Recombination and Increase Efficiency, \href{\doibase 10.1103/PhysRevLett.104.207701}{Phys. Rev. Lett. \textbf{104}, 207701 (2010)}.	

\bibitem{scully2011}
M. O. Scully, K. R. Chapin, K. E. Dorfman, M. B. Kim, and A. Svidzinsky, Quantum heat engine power can be increased by noise-induced coherence, \href{\doibase 10.1073/pnas.1110234108}{Proc. Natl. Acad. Sci. USA \textbf{108}, 15097 (2011)}.

\bibitem{harbola2012}
U. Harbola, S. Rahav, and S. Mukamel, Quantum heat engines: A thermodynamic analysis of power and efficiency, \href{\doibase 10.1209/0295-5075/99/50005}{Europhys. Lett. \textbf{99}, 50005 (2012)}.	

\bibitem{brandner2015}
K. Brandner, M. Bauer, M. T. Schmid, and U. Seifert, Coherence-enhanced efficiency of feedback-driven quantum engines, \href{\doibase 10.1088/1367-2630/17/6/065006}{New J. Phys. \textbf{17}, 065006 (2015)}.

\bibitem{karimi2016}
B. Karimi and J. P. Pekola, Otto refrigerator based on a superconducting qubit: Classical and quantum performance, \href{https://doi.org/10.1103/PhysRevB.94.184503}{Phys. Rev. B {\bf 94}, 184503 (2016)}.

\bibitem{brandner2016}
K. Brandner and U. Seifert, Periodic thermodynamics of open quantum systems, \href{\doibase 10.1103/PhysRevE.93.062134}{Phys. Rev. E \textbf{93}, 062134 (2016)}.	

\bibitem{chen2017}
F. Chen, Y. Gao, and M. Galperin, Molecular Heat Engines: Quantum Coherence Effects, \href{\doibase 10.3390/e19090472}{Entropy \textbf{19}, 472 (2017)}.	 

\bibitem{hardal2017} A. \"{U}. C. Hardal, N. Aslan, C. M. Wilson, and \"{O}. E. M\"{u}stecapl{\i}o\u{g}lu, Quantum heat engine with coupled superconducting resonators, \href{\doibase 10.1103/PhysRevE.96.062120}{Phys. Rev. E \textbf{96}, 062120 (2017)}.

\bibitem{roulet2017}
A. Roulet, S. Nimmrichter, J. M. Arrazola, S. Seah, and V. Scarani, Autonomous rotor heat engine, \href{\doibase 10.1103/PhysRevE.95.062131}{Phys. Rev. E \textbf{95}, 062131 (2017)}.		 	
	   	
\bibitem{brandner2017}
K. Brandner, M. Bauer, and U. Seifert, Universal Coherence-Induced Power Losses of Quantum Heat Engines in Linear Response, \href{\doibase 10.1103/PhysRevLett.119.170602}{Phys. Rev. Lett. \textbf{119}, 170602 (2017)}.

\bibitem{bengtsson2018} J. Bengtsson, M. Nilsson Tengstrand, A. Wacker, P. Samuelsson, M. Ueda, H. Linke, and S. M. Reimann, Quantum Szilard Engine with Attractively Interacting Bosons, \href{\doibase 10.1103/PhysRevLett.120.100601}{Phys. Rev. Lett. \textbf{120}, 100601 (2018)}. 

\bibitem{gonzalez2019}
J. O. Gonz\'{a}lez, J. P. Palao, D. Alonso, and L. A. Correa, Classical emulation of quantum-coherent thermal machines, \href{https://doi.org/10.1103/PhysRevE.99.062102}{Phys. Rev. E {\bf 99}, 062102 (2019)}.

\bibitem{holubec2019}
V. Holubec and T. Novotn\'{y}, Effects of noise-induced coherence on the fluctuations of current in quantum absorption refrigerators, \href{https://doi.org/10.1063/1.5096275}{J. Chem. Phys. {\bf 151}, 044108 (2019)}.

\bibitem{klatzow2019}
J. Klatzow, J. N. Becker, P. M. Ledingham, C. Weinzetl, K. T. Kaczmarek, D. J. Saunders, J. Nunn, I. A. Walmsley, R. Uzdin, and E. Poem, Experimental Demonstration of Quantum Effects in the Operation of Microscopic Heat Engines, \href{\doibase 10.1103/PhysRevLett.122.110601}{Phys. Rev. Lett. \textbf{122}, 110601 (2019)}. 

\bibitem{ptaszynski2018} K. Ptaszy\'{n}ski, Coherence-enhanced constancy of a quantum thermoelectric generator, \href{https://doi.org/10.1103/PhysRevB.98.085425}{Phys. Rev. B {\bf 98}, 085425 (2018)}.

\bibitem{rignon2021}
A. Rignon-Bret, G. Guarnieri, J. Goold, and M. T. Mitchison, Thermodynamics of precision in quantum nanomachines, \href{https://doi.org/10.1103/PhysRevE.103.012133}{Phys. Rev. E \textbf{103}, 012133 (2021)}.

\bibitem{liu2021}
J. Liu and D. Segal, Coherences and the thermodynamic uncertainty relation: Insights from quantum absorption refrigerators, \href{https://doi.org/10.1103/PhysRevE.103.032138}{Phys. Rev. E \textbf{103}, 032138 (2021)}.

\bibitem{myers2021}
N. M Myers, O. Abah, and S. Deffner, Quantum Otto engines at relativistic energies, \href{\doibase 10.1088/1367-2630/ac2756}{New J. Phys. \textbf{23}, 105001 (2021)}.

\bibitem{mu2017}
A. Mu, B. K. Agarwalla, G. Schaller, and D. Segal, Qubit absorption refrigerator at strong coupling, \href{\doibase 10.1088/1367-2630/aa9b75}{New J. Phys. \textbf{19}, 123034 (2017)}.

\bibitem{verley2014a} G. Verley, M. Esposito, T. Willaert, and C. Van den Broeck, The unlikely Carnot efficiency, \href{https://doi.org/10.1038/ncomms5721}{Nat. Commun. {\bf 5}, 4721 (2014)}.

\bibitem{verley2014b} G. Verley, T. Willaert, C. Van den Broeck, and M. Esposito, Universal theory of efficiency fluctuations, \href{https://doi.org/10.1103/PhysRevE.90.052145}{Phys. Rev. E {\bf 90}, 052145 (2014)}.

\bibitem{proesmans2015} K. Proesmans and C. Van den Broeck, Stochastic efficiency: five case studies, \href{https://doi.org/10.1088/1367-2630/17/6/065004}{New. J. Phys. {\bf 17}, 065004 (2015)}.

\bibitem{park2016} J.-M. Park, H.-M. Chun, and J. D. Noh, Efficiency at maximum power and efficiency fluctuations in a linear Brownian heat-engine model, \href{https://doi.org/10.1103/PhysRevE.94.012127}{Phys. Rev. E {\bf 94}, 012127 (2016)}.

\bibitem{holubec2017} V. Holubec and A. Ryabov, Work and power fluctuations in a critical heat engine, \href{https://doi.org/10.1103/PhysRevE.96.030102}{Phys. Rev. E {\bf 96}, 030102(R) (2017)}.

\bibitem{holubec2018} V. Holubec and A. Ryabov, Cycling Tames Power Fluctuations near Optimum Efficiency, \href{https://doi.org/10.1103/PhysRevLett.121.120601}{Phys. Rev. Lett. {\bf 121}, 120601 (2018)}.

\bibitem{manikandan2019} S.~K. Manikandan, L. Dabelow, R. Eichhorn, and S. Krishnamurthy, Efficiency Fluctuations in Microscopic Machines, \href{https://doi.org/10.1103/PhysRevLett.122.140601}{Phys. Rev. Lett. {\bf 122}, 140601 (2019)}.

\bibitem{ito2019} K. Ito, C. Jiang, and G. Watanabe, Universal Bounds for Fluctuations in Small Heat Engines, \href{https://arxiv.org/abs/1910.08096}{arXiv:1910.08096 (2019)}.

\bibitem{vroylandt2020} H. Vroylandt, M. Esposito, and G. Verley, Efficiency Fluctuations of Stochastic Machines Undergoing a Phase Transition, \href{https://doi.org/10.1103/PhysRevLett.124.250603}{Phys. Rev. Lett. {\bf 124}, 250603 (2020)}.

\bibitem{saryal2021} S. Saryal, M. Gerry, I. Khait, D. Segal, and B. K. Agarwalla, Universal Bounds on Fluctuations in Continuous Thermal Machines, \href{https://doi.org/10.1103/PhysRevLett.127.190603}{Phys. Rev. Lett. {\bf 127}, 190603 (2021)}.

\bibitem{holubec2021} V. Holubec and A. Ryabov, Fluctuations in heat engines, \href{https://doi.org/10.1088/1751-8121/ac3aac}{J. Phys. A: Math. Theor. {\bf 55}, 013001 (2021)}.

\bibitem{xu2021}
G.-H. Xu and G. Watanabe, Correlation-enhanced Stability of Microscopic Cyclic Heat Engines, \href{\doibase 10.48550/arXiv.2111.09508}{arXiv:2111.09508 (2021)}.

\bibitem{watanabe2022}
G. Watanabe and Y. Minami, Finite-time thermodynamics of fluctuations in microscopic heat engines, \href{https://doi.org/10.1103/PhysRevResearch.4.L012008}{Phys. Rev. Res. {\bf 4}, L012008 (2022)}.

\bibitem{sinitsyn11} 
N. A. Sinitsyn, Fluctuation relation for heat engines, \href{https://doi.org/10.1088/1751-8113/44/40/405001}{J. Phys. A: Math. Theor. {\bf 44}, 405001 (2011)}.

\bibitem{campisi2014}
M. Campisi, Fluctuation relation for quantum heat engines and refrigerators, \href{https://doi.org/10.1088/1751-8113/47/24/245001}{J. Phys. A: Math. Theor. {\bf 47}, 245001 (2014)}.

\bibitem{holubec2014} V. Holubec, An exactly solvable model of a stochastic heat engine: optimization of power, power fluctuations and efficiency, \href{https://doi.org/10.1088/1742-5468/2014/05/P05022}{J. Stat. Mech., P05022 (2014)}.

\bibitem{pietzonka2018} P. Pietzonka and U. Seifert, Universal Trade-Off between Power, Efficiency, and Constancy in Steady-State Heat Engines, \href{https://doi.org/10.1103/PhysRevLett.120.190602}{Phys. Rev. Lett. {\bf 120}, 190602 (2018)}.

\bibitem{rossnagel2016}
J. Ro{\ss}nagel, S. T. Dawkins, K. N. Tolazzi, O. Abah, E. Lutz, F. Schmidt-Kaler, and K. Singer, A single-atom heat engine, \href{\doibase 10.1126/science.aad6320}{Science {\bf 352}, 325 (2016)}.

\bibitem{lindenfels2019}
D. von Lindenfels, O. Gr\"{a}b, C. T. Schmiegelow, V. Kaushal, J. Schulz, M. T. Mitchison, J. Goold, F. Schmidt-Kaler, and U. G. Poschinger, Spin Heat Engine Coupled to a Harmonic-Oscillator Flywheel, \href{\doibase 10.1103/PhysRevLett.123.080602}{Phys. Rev. Lett. {\bf 123}, 080602 (2019)}.

\bibitem{peterson2019}
J. P. S. Peterson, T. B. Batalh\~{a}o, M. Herrera, A. M. Souza, R. S. Sarthour, I. S. Oliveira, and R. M. Serra, Experimental Characterization of a Spin Quantum Heat Engine, \href{\doibase 10.1103/PhysRevLett.123.240601}{Phys. Rev. Lett. \textbf{123}, 240601 (2019)}.

\bibitem{alhambra2019}
\'{A}. M. Alhambra, M. Lostaglio, and C. Perry, Heat-Bath Algorithmic Cooling with optimal thermalization strategies, \href{\doibase 10.22331/q-2019-09-23-188}{Quantum \textbf{3}, 188 (2019)}.

\bibitem{campisi2022}
A. Solfanelli, A. Santini, and M. Campisi, Quantum thermodynamic methods to refrigerate a qubit on a quantum processing unit, \href{\doibase 10.48550/arXiv.2201.13319}{arXiv:2201.13319 (2022)}.

\bibitem{breuer2002}
H.-P. Breuer and F. Petruccione, \textit{The Theory of Open Quantum Systems} (Oxford University Press, Oxford, 2002).

\bibitem{rivas2014}
\'{A}. Rivas, S. F. Huelga, and M. B. Plenio, Quantum non-Markovianity: characterization, quantification and detection, \href{https://doi.org/10.1088/0034-4885/77/9/094001}{Rep. Prog. Phys. \textbf{77}, 094001 (2014)}.

\bibitem{wenderoth2021}
S. Wenderoth, H.-P. Breuer, and M. Thoss, Non-Markovian effects in the spin-boson model at zero temperature, \href{\doibase 10.1103/PhysRevA.104.012213}{Phys. Rev. A \textbf{104}, 012213 (2021)}.

\bibitem{vasile2011}
R. Vasile, S. Olivares, M. G. A. Paris, and S. Maniscalco, Continuous-variable quantum key distribution in non-Markovian channels, \href{\doibase 10.1103/PhysRevA.83.042321}{Phys. Rev. A \textbf{83}, 042321 (2011)}.

\bibitem{bylicka2014}
B. Bylicka, D. Chru\'{s}ci\'{n}ski, and S. Maniscalco, Non-Markovianity and reservoir memory of quantum channels: a quantum information theory perspective, \href{\doibase 10.1038/srep05720}{Sci. Rep \textbf{4}, 5720 (2014)}.

\bibitem{chin2012}
A. W. Chin, S. F. Huelga, and M. B. Plenio, Quantum Metrology in Non-Markovian Environments, \href{\doibase 10.1103/PhysRevLett.109.233601}{Phys. Rev. Lett. \textbf{109}, 233601 (2012)}.

\bibitem{zhang2017}
W.-Z. Zhang, Y. Han, B. Xiong, and L. Zhou, Optomechanical force sensor in a non-Markovian regime, \href{\doibase 10.1088/1367-2630/aa68d9}{New J. Phys. \textbf{19}, 083022 (2017)}.

\bibitem{huelga2012}
S. F. Huelga, \'{A}. Rivas, and M. B. Plenio, Non-Markovianity-Assisted Steady State Entanglement, \href{\doibase 10.1103/PhysRevLett.108.160402}{Phys. Rev. Lett. \textbf{108}, 160402 (2012)}.

\bibitem{heineken2021}
D. Heineken, K. Beyer, K. Luoma, and W. T. Strunz, Quantum-memory-enhanced dissipative entanglement creation in nonequilibrium steady states, \href{\doibase 10.1103/PhysRevA.104.052426}{Phys. Rev. A \textbf{104}, 052426 (2021)}.

\bibitem{spaventa2022}
G. Spaventa, S. F. Huelga, and M. B. Plenio, Capacity of non-Markovianity to boost the efficiency of molecular switches, \href{\doibase 10.1103/PhysRevA.105.012420}{Phys. Rev. A \textbf{105}, 012420 (2022)}.

\bibitem{zhang2014}
X. Y. Zhang, X. L. Huang, and X. X. Yi, Quantum Otto heat engine with a non-Markovian reservoir, \href{\doibase 10.1088/1751-8113/47/45/455002}{J. Phys. A: Math. Theor. \textbf{47}, 455002 (2014)}.

\bibitem{thomas2018}
G. Thomas, N. Siddharth, S. Banerjee, and S. Ghosh, Thermodynamics of non-Markovian reservoirs and heat engines, \href{\doibase 10.1103/PhysRevE.97.062108}{Phys. Rev. E \textbf{97}, 062108 (2018)}.

\bibitem{wiedmann2020}
M. Wiedmann, J. T. Stockburger, and J. Ankerhold, Non-Markovian dynamics of a quantum heat engine: out-of-equilibrium operation and thermal coupling control, \href{\doibase 10.1088/1367-2630/ab725a}{New J. Phys. \textbf{22}, 033007 (2020)}.

\bibitem{shirai2021}
Y. Shirai, K. Hashimoto, R. Tezuka, C. Uchiyama, and N. Hatano, Non-Markovian effect on quantum Otto engine: Role of system-reservoir interaction, \href{\doibase 10.1103/PhysRevResearch.3.023078}{Phys. Rev. Research \textbf{3}, 023078 (2021)}.

\bibitem{abiuso2019}
P. Abiuso and V. Giovannetti, Non-Markov enhancement of maximum power for quantum thermal machines, \href{\doibase 10.1103/PhysRevA.99.052106}{Phys. Rev. A \textbf{99}, 052106 (2019)}.

\bibitem{camati2020}
P. A. Camati, J. F. G. Santos, and R. M. Serra, Employing non-Markovian effects to improve the performance of a quantum Otto refrigerator, \href{\doibase 10.1103/PhysRevA.102.012217}{Phys. Rev. A \textbf{102}, 012217 (2020)}.

\bibitem{horodecki2013}
M. Horodecki and J. Oppenheim, Fundamental limitations for quantum and nanoscale thermodynamics, \href{https://doi.org/10.1038/ncomms3059}{Nat. Commun. \textbf{4}, 2059 (2013)}.

\bibitem{brandao2013}
F. G. S. L. Brand\~{a}o, M. Horodecki, J. Oppenheim, J. M. Renes, and R. W. Spekkens, Resource Theory of Quantum States Out of Thermal Equilibrium, \href{https://doi.org/10.1103/PhysRevLett.111.250404}{Phys. Rev. Lett. \textbf{111}, 250404 (2013)}.

\bibitem{brandao2015}
F. G. S. L. Brand\~{a}o, M. Horodecki, N. H. Y. Ng, J. Oppenheim, and S. Wehner, The second laws of quantum thermodynamics, \href{https://doi.org/10.1073/pnas.1411728112}{Proc. Natl. Acad. Sci. U.S.A. \textbf{112}, 3275 (2015)}.

\bibitem{goold2016}
J. Goold, M. Huber, A. Riera, L. del Rio,
and P. Skrzypczyk, The role of quantum information in thermodynamics---a topical review, \href{\doibase 10.1088/1751-8113/49/14/143001}{J. Phys. A: Math. Theor. \textbf{49}, 143001 (2016)}.

\bibitem{lostaglio2019}
M. Lostaglio, An introductory review of the resource theory approach to thermodynamics, \href{\doibase 10.1088/1361-6633/ab46e5}{Rep. Prog. Phys. \textbf{82}, 114001 (2019)}.

\bibitem{feldmann2000}
T. Feldmann and R. Kosloff, Performance of discrete heat engines and heat pumps in finite time, \href{\doibase 10.1103/PhysRevE.61.4774}{Phys. Rev. A \textbf{61}, 4774 (2000)}.

\bibitem{lobejko2020}
M. \L{}obejko, P. Mazurek, and M. Horodecki, Thermodynamics of Minimal Coupling Quantum Heat Engines, \href{\doibase 10.22331/q-2020-12-23-375}{Quantum \textbf{4}, 375 (2020)}.

\bibitem{lostaglio2021}
M. Lostaglio and K. Korzekwa, Continuous thermomajorization and a complete set of laws for Markovian thermal processes, \href{\doibase 10.48550/arXiv.2111.12130}{arXiv:2111.12130 (2021)}.

\bibitem{cwiklinski2015}
P. \'{C}wikli\'{n}ski, M. Studzi\'{n}ski, M. Horodecki, and J. Oppenheim, Limitations on the Evolution of Quantum Coherences: Towards Fully Quantum Second Laws of Thermodynamics, \href{\doibase 10.1103/PhysRevLett.115.210403}{Phys. Rev. Lett. \textbf{115}, 210403 (2015)}.

\bibitem{lostaglio2018}
M. Lostaglio, \'{A}. M. Alhambra, and C. Perry, Elementary Thermal Operations, \href{\doibase 10.22331/q-2018-02-08-52}{Quantum \textbf{2}, 52 (2018)}.

\bibitem{touchette2009}
H. Touchette, The large deviation approach to statistical mechanics, \href{\doibase 10.1016/j.physrep.2009.05.002}{Phys. Rep. \textbf{478}, 1 (2009)}.

\bibitem{chiuchiu2018}
D. Chiuchi\`{u} and S. Pigolotti, Mapping of uncertainty relations between continuous and discrete time, \href{\doibase 10.1103/PhysRevE.97.032109}{Phys. Rev. E \textbf{97}, 032109 (2018)}.

\bibitem{naderi2005}
M. H. Naderi, M. Soltanolkotabi, and R. Roknizadeh, A theoretical scheme for generation of nonlinear coherent states in a micromaser under intensity-dependent Jaynes-Cummings model, \href{https://doi.org/10.1140/epjd/e2004-00197-8}{Eur. Phys. J. D {\bf 32}, 397 (2005)}.

\bibitem{aberg2014}
J. {\AA}berg, Catalytic Coherence, \href{https://doi.org/10.1103/PhysRevLett.113.150402}{Phys. Rev. Lett. {\bf 113}, 150402 (2014)}.

\bibitem{bera2022}
M. L. Bera, S. Juli\`{a}-Farr\'{e}, M. Lewenstein, and M. N. Bera, Quantum heat engines with Carnot efficiency at maximum power, \href{https://doi.org/10.1103/PhysRevResearch.4.013157}{Phys. Rev. Research {\bf 4}, 013157 (2022)}.

\bibitem{guthrie2021}
A. Guthrie, C. D. Satrya, Y.-C. Chang, P. Menczel, F. Nori, and J. P. Pekola, A Cooper-Pair Box Architecture for Cyclic Quantum Heat Engines, \href{https://doi.org/10.48550/arXiv.2109.03023}{arXiv:2109.03023 (2021)}.

\bibitem{ronzani2018}
A. Ronzani, B. Karimi, J. Senior, Y.-C. Chang, J. T. Peltonen, C. Chen, and J. P. Pekola, Tunable photonic heat transport in a quantum heat valve, \href{https://doi.org/10.1038/s41567-018-0199-4}{Nat. Phys. {\bf 14}, 991 (2018)}.

\bibitem{senior2020}
J. Senior, A. Gubaydullin, B. Karimi, J. T. Peltonen, J. Ankerhold, and J. P. Pekola, Heat rectification via a superconducting artificial atom, \href{https://doi.org/10.1038/s42005-020-0307-5}{Commun. Phys. {\bf 3}, 40 (2020)}.

\end{thebibliography}
\end{document}